\documentstyle[12pt]{article}
\begin{document}
\baselineskip 20pt
\begin{titlepage}
\begin{flushright}{FAU-TP3-97/2}
\end{flushright}
\vskip 3.0cm
\begin{center}
{\Large {\bf QCD at Finite Extension }}
\vskip 0.5cm
F. Lenz and M. Thies
\vskip 0.2cm
{\it Institute for Theoretical Physics III, University of
Erlangen-N\"urnberg, \\
Staudtstr. 7, 91058 Erlangen, Germany}
\end{center}
\vskip 3.0cm
\begin{abstract}
A study of QCD at finite extension is presented  and the relation to QCD
at finite
temperature and in the infinite momentum frame is discussed.
The dynamics of Polyakov loops is investigated and shown to be described
by functional integrals with finite range of integration.
Consequences of this non-Gaussian  form of the generating functional
concerning the
QCD vacuum, gluonic excitations and  confinement are discussed.
\end{abstract}

\end{titlepage}

The study of QCD at finite extension, i.e., in a geometry where the system
is of finite extent ($L_{3}$) in one direction
($x_{3}$), is of interest for various reasons. First, with
$L_{3}$ a parameter is introduced which helps to control infrared
ambiguities. This is of particular importance when using axial like
gauges which are appropriate for such a geometry
(for an early discussion of ambiguities in the axial gauge, see
Ref. \cite{Schwinger}).

Second, by covariance, QCD at finite extension is equivalent to finite
temperature QCD. This equivalence of finite extension and finite
temperature of relativistic field theories has been noted in Ref. \cite{Toms}
 and used e.g. in a discussion of the finite
temperature quark propagators \cite{Koch}.
By rotational invariance in the Euclidean, the value of the
partition function of a system with finite extension $L_{3}$
in 3 direction and $\beta$  in 0 direction is invariant under the exchange
of  these two extensions,
\begin{equation}
Z\left(\beta,L_{3}\right)=Z\left(L_{3},\beta\right) \ ,
\label{FE1}
\end{equation}
provided bosonic (fermionic) fields satisfy periodic (antiperiodic)
boundary conditions in both time
and 3 coordinate.
As a consequence of (1), energy density and pressure are related
by
\begin{equation}
\epsilon\left(\beta,L_{3}\right)=-p\left(L_{3},\beta\right) \ .
\label{FE2}
\end{equation}
For a system of non-interacting particles this relation connects energy
density or pressure of the Stefan Boltzmann law with the corresponding
quantities measured in the Casimir effect.

In QCD, covariance also implies by Eq. (2)
that at zero temperature a confinement-deconfinement
transition occurs when compressing
the QCD vacuum (i.e. decreasing $ L_{3}$).
From lattice gauge calculations \cite{Kanayo} we infer
that this transition occurs
at a critical extension $L^{c}_{3} \approx 0.8$ fm in the absence of
quarks and at $L^{c}_{3} \approx 1.3$ fm when quarks are included.
For extensions smaller than $L_3^c$, the energy density
and pressure reach values which are typically 80 \% of the
corresponding  ``Casimir" energy and pressure. The order parameter
which characterizes the phases of QCD and in particular the realization
of the center symmetry (in the absence of quarks)
is the vacuum expectation
value  of the Polyakov loop operator at finite
temperature \cite{Svetitsky} and correspondingly
of the operator
\begin{equation}
W \left(x_{\perp}\right) = N_c^{-1} \mbox{tr}\,
P\exp\left\{ig \int_{0}^{L_{3}} dx_{3}  A_{3} \left(x\right)  \right\}
\label{FE3}
\end{equation}
at finite extension ($x_{\perp} = (x_{0}, x_{1}, x_{2})$).

When compressing the system beyond the typical length scales of strong
interaction physics (e.g. beyond a typical hadron radius $R$),
two further limits are of interest. Again as supported by
results from lattice calculations \cite{Reisz}, correlation
functions at transverse
momenta or energies $|p| \ll 1/L_{3} $ are dominated by the
zero ``Matsubara wave-numbers" in 3-direction and can therefore be
expected to be given by the dimensionally reduced QCD$_{2+1}$.
On the other hand we may consider excitations with large momenta $p$ in
the 3 direction such that the Lorentz-contracted extension of a
hadron (of mass $m$) is small in comparison to $L_{3}$,
\begin{equation}
mR/p \ll L_{3} \ll R \ .
\label{FE4}
\end{equation}
For sufficiently high momenta, spectrum and structure of hadrons
should not be
affected by the finite extension, although the vacuum on
which these excitations are built is not the confining ground
state with broken chiral symmetry,  but that of the
quark gluon plasma with the chiral symmetry restored.
 This situation of large momentum
excitations at small extension is closely related to the light cone
or infinite
momentum frame limit in which, because of the off-diagonal metric,
a finite extension along the light-cone
space axis $x^{-}$ actually describes an interval of vanishing
invariant length \cite{Lenz1}.

For the theoretical treatment of QCD at finite extension an axial
type gauge is particularly appropriate. Within the canonical formalism,
the Gauss law can be resolved explicitly \cite{Lenz2}. The result for the
SU(2) Yang Mills theory is summarized in the generating
functional
\begin{equation}
Z\left[J,j_{3}\right] = \int D\left[A_{\mu}\right] d\left[a_{3}\right]
\exp \left\{
  i S_{\rm YM} \left[A_{\mu},a_{3}\right] + i S_{\rm gf}
\left[A_{\mu}^{3}\right]
+ i\int d^{4}x\left(A^{a}_{\mu}  J^{a,\mu} + a_{3}  j^{3} \right) \right\}
\ ,
\label{FE5}
\end{equation}
where in the standard Yang Mills action the 3 component of the
gluon field $A_{3} (x) $ is replaced by the neutral (i.e. diagonal)
2-dimensional field
$a_{3} (x_{\perp}) $ (cf. also Ref. \cite{Reinh}).
At finite extension $L_{3} $, this two dimensional
field can be eliminated neither in QED nor in QCD. In QED it
describes transverse photons polarized in 3 direction and propagating
in the 1-2 plane, while in SU(2) QCD, $a_{3} (x_{\perp})$ are the eigenvalues
of the (untraced) ``Polyakov loop" variable $W(x_{\perp})$ of Eq. (3).
After an appropriate choice of coordinates in color space,
$W(x_{\perp})$ can be written as
\begin{equation}
W \left(x_{\perp}\right)= \cos\left(g L_{3} a_{3}\left(x_{\perp}
\right)/2  \right) \ .
\label{FE6}
\end{equation}
The presence of the two dimensional field $a_{3} (x_{\perp})$ makes a
further gauge fixing of the transverse ($\mu = 0, 1, 2$), neutral, $x_{3}$
independent gauge fields necessary, which is achieved by the additional
contribution to the action in Eq. (\ref{FE5}). The following choice of
the gauge fixing
term is particularly convenient for perturbative calculations,
\begin{equation}
S_{\rm gf} \left[A_{\mu}^{3}\right] =- \frac{1}{2 L_{3}^{2}}
\left(\int_{0}^{L_{3}}dx_{3}\partial ^{\mu}A^{3}_{\mu}
\left(x\right)\right)^{2} .
\label{FE7}
\end{equation}
The integration measure for the $a_{3}$
functional integral is given by
\begin{equation}
d\left[a_{3}\right] = \prod_{y_{\perp}} \sin^{2}\left( gL_{3}
a_{3}(y_{\perp})/2 \right)
\Theta \left( a_3(y_{\perp}) \right) \Theta \left( 2\pi/g L_3
-a_3(y_{\perp}) \right)
 da_{3}\left(y_{\perp}\right)
\label{FE8}
\end{equation}
and accounts for the fact that $gL_3a_3/2$
appears in the parametrization of the group manifold $S_{3} $ as
the first polar angle.
As a consequence the corresponding part of the Haar measure enters
 with its finite range of integration.  In the canonical formulation,
this Faddeev Popov determinant arises as a Jacobian modifying the
kinetic energy of the Polyakov loop variables $a_{3} (x_{\perp})$
\cite{Lenz2}.
In QED the same procedure yields the standard flat measure for
$a_{3} (x_{\perp})$.

In most of the approaches to finite temperature QCD in the
temporal gauge, one ignores the
finite
range of integration and accounts for the factor $\sin^{2}
\left(g L_{3} a_{3}/2\right) $ in the volume element
by introducing ghost fields $c_{a}\left(x\right)$,
\begin{equation}
\prod_{y_{\perp}} \sin^{2}
\left( gL_{3}a_{3}(y_{\perp}) /2 \right)
 =  \int d[c] d[c^{\dagger}]
\exp\left\{i \int d^{4}x c^{\dagger}_{a}
\left(x\right) \left(-i \partial _{3}
\delta^{ab}+ ig \epsilon^{ab3} a_{3}^{c}
\left(x_{\perp}\right)\right)c_{b} \left(x\right) \right\} .
\label{FE9}
\end{equation}
In perturbative treatments, such ghosts have no effect, as is
most easily seen in dimensional regularization,
and therefore the QCD measure is effectively replaced
by the  flat measure appropriate for QED. Debye screening
is obtained as a result of this procedure but also, as in
QED, a shift symmetry (Z) of the effective potential for the
Polyakov loop variables $a_{3}$  not present in the original theory
\cite{Weiss}.

We have not been able to justify the above procedure. We find that
the functional integral over $a_{3}$ cannot be approximated by a Gaussian
integral with corresponding perturbative corrections. Rather, the
relevant limit is that of a vanishing instead of an infinite range of
integration, as is easily seen when considering the propagator
for non-interacting Polyakov loop variables $W(x_{\bot})$
in the discretized
Euclidean formulation (with lattice spacing $\ell$ and
unit vectors ${\bf e}$),
\begin{eqnarray}
\langle \Omega|T\left[ W(x_{\bot}) W(0)
  \right]|\Omega \rangle
& = &  \frac{1}{4}
\delta_{\bf x, 0}
+ \mbox{O} \left( \frac{\ell}{g^2 L_3}\right) \left(
\delta_{\bf x,0}+ \sum_{\pm {\bf e}} \delta_{\bf x,e}
\right)
\nonumber \\
& \approx & \frac{\ell^{3}}
{4}\delta^{ \left(3\right)} \left(x_{\perp}\right) \ .
\label{FE10}
\end{eqnarray}
Hopping terms are suppressed by powers of $ \ell /g^2 L_3$, and the
propagator becomes
local in the continuum limit.
We note that this property is due to the ``macroscopic" nature
of the variable $a_{3}(x_{\perp}) $. When describing a single link
variable on the lattice, the factor $\ell/L_{3}$ does not appear and the
discretized form of the free propagator is obtained. The
winding of $a_{3}$ around
the circle in 3 direction apparently provides an infinite
inertia. Propagation of excitations induced by $a_{3} (x_{\perp})$ can
consequently only arise by coupling to the other microscopic degrees
of freedom. Formally this suggests the Polyakov loop variables
$a_{3} $ to
be integrated out  by disregarding the contribution of the free $a_{3}$
action, but keeping the coupling to the other degrees of freedom.
Also disregarding possible effects from singular field configurations,
the following effective action is obtained
\begin{equation}
S_{\rm eff} \left[A_{\mu}\right] = S_{\rm YM} \left[A_{\mu}, A_{3}=0\right]
+ S_{\rm gf} \left[A_{\mu}^{3}\right] +  M^{2}/2 \sum_{a=1,2}
 \int d^{4}x A^{a}_{\mu} \left(x\right)  A^{a,\mu} \left(x\right) .
\label{FE11}
\end{equation}
In this effective action,  the Polyakov loop variables have left
their signature in the
geometrical mass term of the charged gluons
\begin{equation}
M^{2}=\left(\pi^{2}/3-2\right)/L_{3}^{2}
\label{FE12}
\end{equation}
and in the  change to antiperiodic  boundary conditions
\begin{equation}
A_{\mu}^{1,2} \left(x_{\perp}, L_{3}\right) = - A_{\mu}^{1,2}
\left(x_{\perp}, 0\right) \ ,
\label{FE12a}
\end{equation}
while the neutral gluons remain massless and periodic.
The antiperiodic boundary conditions reflect the mean value of the Polyakov
loop variables, the geometrical mass their fluctuations;
notice that in both of these corrections, the coupling constant has
dropped out.
The role of the order parameter is taken over by the neutral color
current in 3-direction,
\begin{equation}
u \left(x_{\perp}\right) = \int_{0}^{L_{3}} dx_{3}
\epsilon^{ab3} A^{a}_{\mu} \left(x\right)
\partial_3  A^{b,\mu} \left(x\right)  \ ,
\label{FE13}
\end{equation}
which determines the vacuum expectation value of the Polyakov loops
\begin{equation}
\langle \Omega| W \left(x_{\perp}\right)|\Omega \rangle \quad
\propto \quad \langle \Omega| u \left(x_{\perp}\right)|\Omega \rangle
\label{FE14}
\end{equation}
and the corresponding correlation function
\begin{equation}
\langle \Omega| T\left[W \left(x_{\perp}\right) W \left(0\right)\right]
|\Omega \rangle  \quad  \propto \quad \langle \Omega| T\left[u
\left(x_{\perp}\right) u \left(0\right)\right] |\Omega \rangle \ .
\label{FE15}
\end{equation}
Formulated  in these variables, the center symmetry acts as
charge conjungation \cite{foot1}
\begin{equation}
C:\quad \quad A_{\mu}^{3} \rightarrow
- A_{\mu}^{3}\quad , \quad A_{\mu}^{1}+i A_{\mu}^{2}
\rightarrow A_{\mu}^{1}-i A_{\mu}^{2} \ .
\label{FE16}
\end{equation}

The treatment of the Polyakov loop variables required by the finite
range of integration leads to a picture of QCD which is quite
different from that based on the treatment of the $a_{3}$ functional
integral as an approximate Gaussian one. Among the most important
differences, we note that the perturbative vacuum corresponding
to the effective action (\ref{FE11}) is an eigenstate of the center symmetry.
In the perturbative ground state, no color currents are present, hence
\begin{equation}
\langle\Omega_{\rm pt}| W \left(x_{\perp}\right)|\Omega_{\rm pt}
\rangle = 0  \ .
\label{FE17}
\end{equation}
Loosely speaking we can take this as an indication for an infinite
free energy of a static quark. A more precise
characterization  of the realization of confinement
can be obtained from properties of the associated correlation
function (\ref{FE15}) which, in the Euclidean, determines the
static quark-antiquark
interaction energy \cite{Svetitsky}. Due to the locality
property (\ref{FE10}), this static quark-antiquark
potential is given directly by the $a=b=3, \mu = \nu = 3$ component
of the  vacuum polarization tensor $\Pi_{\mu \nu}^{ab} $
and not by the zero mass propagator
of the Gaussian approximation to the Polyakov loop action.
Up to an irrelevant factor
we have in the Euclidean
%################################################################
\input FEYNMAN
\vskip -0.5 in
\begin{picture}(68000,8000)
\drawloop\gluon[\E 0](30000,0)
\Xfour=\loopmidx
\Yfour=\loopmidy
\advance\Xfour by -28000
\advance\Yfour by -300
\put(\Xfour,\Yfour){$ \exp{\left\{-L_{3}V\left(r \right)\right\}}=
\langle \Omega|T \left[u ( x^{E}_{\perp}) u\left(0\right)\right]|
\Omega\rangle
 = $}

\advance\loopmidx by -2650
\put(\loopmidx,\loopmidy){\circle*{550}}
\advance\loopmidx by -1800
\advance\loopmidy by  -200

\put(\loopmidx,\loopmidy){$  x^{E}_{\perp} $}
\advance\loopmidy by  200
\advance\loopmidx by  7100
\put(\loopmidx,\loopmidy){\circle*{550}}
\advance\loopmidx by  500
\advance\loopmidy by  -400
\put(\loopmidx,\loopmidy){$ 0$}
\end{picture}
\vskip -.52in

\begin{equation}
\label{FE17a}
\end{equation}
\vskip .45in
%################################################################
\noindent
This identity implies a linearly rising potential at large distances,
if the vacuum expectation value of the Polyakov loop operator vanishes
and if the spectrum of states
excited by $u$ exhibits a gap $\Delta E$,
\begin{equation}
V\left(r \right) \rightarrow  \sigma r = \Delta E  r/L_3 \ .
\label{FE17b}
\end{equation}
Thus in this axial like gauge, confinement  is connected to a shift
in the spectrum of gluonic excitations to excitation energies
\begin{equation}
 E \geq  \sigma L_{3} \ .
\label{FE17c}
\end{equation}
After a gradual decrease of this threshold with decreasing $L_{3}$,
the whole spectrum of excitations becomes suddenly available when at
the deconfinement transition the string
tension vanishes. At the perturbative level already a linearly rising
potential is obtained, however with a string tension decreasing with
increasing extension
 ($ \propto L_{3}^{-2}$). For small distances,
Coulomb-like behaviour
 requires  this particular component of the vacuum polarization
tensor to possess an
 essential singularity at infinite momentum
\begin{equation}
\int d^{3}x e^{i px}  \langle \Omega|
T\left[u \left(x\right) u
\left(0\right)\right] |\Omega \rangle   \rightarrow
e^{-\sqrt{g^{2}L_{3}p/\pi}}
\label{FE17d}
\end{equation}
which, in the Euclidean, implies a density of intermediate states
contributing  to the vacuum
 polarization which increases exponentially with the square root
of the excitation energy.
 A perturbative evaluation of the vacuum polarization therefore
has to yield
 increasingly high powers of $L_{3}p$ with increasing order \cite{Lenz3}.

 With respect to perturbative treatments, we remark that the
difficulties encountered in finite
temperature perturbation theory \cite{Linde} are not expected to  arise
in this perturbative, confining phase.
With the charged gluons satisfying antiperiodic boundary conditions,
the infrared properties of QCD at finite extension should resemble
those of QED. Moreover, the finite extension gluon
propagator is well defined and, unlike the continuum axial gauge propagator,
does not
need a prescription how to handle spurious double poles \cite{Landshoff}.

A final remark concerns the perturbative phase with its signatures of
confinement. This phase is most likely not relevant for QCD at
extensions smaller than $L_3^c$. Not only do
we expect the center symmetry to be
 broken at small extensions but also dimensional reduction to
 QCD$_{2+1}$ to happen. On the other hand, in the process of dimensional
reduction, charged gluons decouple
from the low-lying excitations due to their antiperiodic boundary conditions;
the small extension or high temperature
limit in this phase is QED$_{2+1}$. Thus in the deconfinement phase
transition arising when compressing the QCD vacuum, a change to periodic
boundary conditions as well as the disappearence of the geometrial
mass must occur. This change in boundary conditions results in a change
in Casimir energy density and pressure which for
non-interacting gluons (and neglecting the effects of the
geometrical mass) is given by
\begin{equation}
\Delta \epsilon=  - \pi^{2} /12 L_3^4 \ ,\qquad \Delta p =
3\Delta \epsilon \ . \label{FE18}
\end{equation}
This estimate is of the order of magnitude of the change in the energy
density across the confinement-deconfinement transition when compressing
the system,
\begin{equation}
\Delta \epsilon=  -0.45 /L_3^4 \ ,
\label{FE19}
\end{equation}
deduced from the finite temperature lattice calculation of Ref.
\cite{Engels}.
If QCD in the high temperature or small extension quark gluon plasma
phase is to be described perturbatively, one has to abandon
the non-perturbative resolution which is implicit to the temporal
or axial like gauge with its characteristic $N-1$ Polyakov loops
as zero modes. Starting point has to be  QCD at $g=0$, and for this
 U(1)$^{N^2-1}$ theory an axial gauge with $N^2-1$ photons as zero modes
is e.g. appropriate. This procedure breaks the center symmetry
and the high temperature limit results \cite{Lenz3}.

In summary we have investigated QCD at finite extension in an axial
like gauge. Novel aspects in the theoretical description have emerged
due to the peculiar dynamics of the
Polyakov loops; these variables are described by functional integrals
which do not have
a natural Gaussian limit. By elimination
of these variables an effective description has been derived. It
yields a QCD ground state which already
at the perturbative level exhibits certain characteristics of the
confining vacuum and permits to connect confinement with properties
of the spectrum of
gluonic excitations.

Details of this investigation, the development of perturbation theory
and its application to the static quark-antiquark potential
as well as generalizations to SU(3) and full QCD
with fermions will be
presented elsewhere \cite{Lenz3}.

We thank V.L. Eletskii and A.C. Kalloniatis for valuable discussions.

\newpage
\bibliographystyle{unsrt}

\end{document}